\documentstyle[12pt,makeidx,epsfig,rotating]{article}
\rm
 
\makeatletter
\renewcommand{\section}{\@startsection{section}{1}{0mm}{0.5cm}{0.2cm}{\bf}}

\makeatother

\def\gsim{\mathrel{\rlap{\raise.4ex\hbox{$>$}} {\lower.6ex\hbox{$\sim$}}}}
\def\lsim{\mathrel{\rlap{\raise.4ex\hbox{$<$}} {\lower.6ex\hbox{$\sim$}}}}

\newcommand{\chha}{$\tilde{\chi}_1^{\pm}$}

\newcommand{\chnb}{$\tilde{\chi}_2^{0}$}

\newcommand{\g}{$\tilde{g}$}
\newcommand{\q}{$\tilde{q}$}

\begin{document}
\begin{titlepage}

\begin{flushright}
{\bf CMS CR 1997/012}
\end{flushright}

\vspace*{1cm}

\begin{center} 
 
{\bf SEARCHES FOR SUSY AT LHC}

\bigskip
\bigskip

   For the CMS Collaboration    \\

\medskip

   Avtandyl Kharchilava \footnote{E-mail: Avto.Kharchilava@cern.ch} \\
 
  {\em Institute of Physics, Georgian Academy of Sciences, Tbilisi \/}

\bigskip
\bigskip
\bigskip
\bigskip
\bigskip
\bigskip
\bigskip
\bigskip
\bigskip
\bigskip
\bigskip
\bigskip
\bigskip
\bigskip
\bigskip
\bigskip
\bigskip
\bigskip
\bigskip
 
\begin{abstract}

One of the main motivations of experiments at the LHC is to
search for SUSY particles. The talk is based on recent analyses,
performed
by CMS Collaboration, within the framework of the Supergravity
motivated minimal SUSY extension of the Standard Model. The
emphasis is put on leptonic channels. The strategies for obtaining
experimental signatures for strongly and weakly interacting sparticles
productions, as well as examples of determination of SUSY masses
and model parameters are discussed. The domain of parameter space where
SUSY can be discovered is investigated.
Results show, that if SUSY is of relevance
at Electro-Weak scale it could hardly escape detection at LHC.

\end{abstract}

\bigskip
\bigskip
\bigskip

Presented at {\it XXXII Rencontres de Moriond: QCD and High Energy
Hadronic Interactions}, Les Arcs, Savoie, France, March 22-29, 1997

\end{center}
\end{titlepage}
 
\section{Introduction}

The Standard Model {\small (SM)}, despite
its phenomenological successes, is most likely a low energy effective
theory of spin-1/2 matter fermions interacting via spin-1
gauge bosons. A good candidate for the new physics
beyond the {\small SM} is the Supersymmetry {\small (SUSY)}. In the
minimal version
it doubles the number of known particles,
introducing scalar (fermion) partners to ordinary fermions (bosons)
with the same couplings. These provide
cancellation of divergences in the radiative
corrections of the SM, but necessarily, the masses of
super-partners should be of the order of
Electro-Weak {\small (EW)} scale, i.e. $\lsim 1$ TeV.

As a framework we have chosen
the minimal Supergravity {\small (mSUGRA)}, which
is the most fully investigated model \cite{baer1}, where only five
extra parameters
need to be specified: the universal scalar ($m_{0}$), gaugino
($m_{1/2}$) masses and trilinear term ($A_{0}$) which are
fixed at the gauge coupling unification scale;
the ratio of the vacuum expectation
values of the two Higgs fields (tan$\beta$) and the sign of the
Higgsino mixing parameter ($sign(\mu)$). Sparticle
masses and couplings at the {\small EW} scale are then evolved
via the renormalization group equations. Obtained {\small SUSY} mass
spectrum is dependent most strongly on $m_{0}$ and $m_{1/2}$.
In the following we limit ourselves to the choice of
tan$\beta=2$, $A_{0}=0$ and $\mu<0$.
Fig.1 shows the isomass contours for gluinos ($\tilde{g}$),
charginos ($\tilde{\chi}^{\pm}$), neutralinos ($\tilde{\chi}^{0}$),
sleptons ($\tilde{l}$) in ($m_{0}$, $m_{1/2}$) parameter space.

In $R$-parity conserving {\small SUSY} models sparticles are
produced in pairs and a stable Lightest Supersymmetric Particle
({\small LSP}; which is the $\tilde{\chi}_1^{0}$ in {\small mSUGRA})
appears at the end
of each sparticle decay chain. It is weakly interacting and escapes
detection thus leading to a classical $E_T^{miss}$ signature.
Due to the escaping {\small LSP}'s the masses of
sparticles cannot be reconstructed explicitly.
Usually, one characterizes the {\small SUSY} signal significance ($S$)
by an excess of events ($N_S$) over the {\small SM} background expectation
($N_B$): $S = N_S / \sqrt{N_S + N_B}$.
In some cases, the background to a
particular {\small SUSY} channel is {\small SUSY} itself.

The goal of the current analysis is to evaluate the domain
of ($m_{0}$, $m_{1/2}$) parameter space in which {\small SUSY}
can be discovered, estimate the reach in gluino, squark, slepton,
chargino, neutralino masses in various channels,
develop methods to determine {\small SUSY} masses and model parameters,
understand the instrumental limiting factors and contribute to 
detector optimization before the design is frozen \cite{lhcc}.

In the following the {\small SM} backgrounds are generated
with P{\footnotesize YTHIA} and {\small mSUGRA} processes with
{\small ISAJET}.
The {\small CMS} detector performances are parameterized on the basis
of detailed simulations \cite{sal1}.

\section{Gluino/Squark Production}

At {\small LHC} energies, the total {\small SUSY} particles production
cross-section
are largely dominated by strongly interacting sparticles.
Thus a typical high mass {\small SUSY} signal has squarks and/or gluinos
which decay through a number of steps to quarks, gluons, charginos,
neutralinos, W, Z, Higgses and ultimately to a
stable $\tilde{\chi}_{1}^{0}$.
For instance, the branching ratios of $\tilde{g}$, $\tilde{q}$ decays
into $\tilde{\chi}_1^{\pm}$ ($\tilde{\chi}_2^{0}$) are
complementary in ($m_{0}$, $m_{1/2}$) parameter space and
exceed 30$\div$50$\%$. Furthermore, at values of
$m_{0}$, $m_{1/2}$ $\gsim$ 200 GeV the decays $\tilde{\chi}_1^{\pm}
\rightarrow$ W$^{\pm}\tilde{\chi}_{1}^{0}$ are dominant,
giving an isolated lepton ($\mu$ or e from W)
in about 20$\%$ of cases.
In the region of $m_{0}$, $m_{1/2}$ $\lsim$ 200 GeV
the leptonic branching ratios are even higher (here, also
B$(\tilde{\chi}_{2}^{0}
\rightarrow l^+l^- + invisible)$ may reach 10$\div$20$\%$
being an additional important source of isolated leptons).
The final state has thus a number of hard jets, missing energy 
(2$\tilde{\chi}_{1}^{0}$ + neutrinos) and a variable number
of leptons, depending on the decay chain.
The {\small SM} backgrounds considered are:
$t \bar{t}$, W+jets, Z+jets, WW, ZZ, ZW, Z$b\bar{b}$,
QCD (2 $\rightarrow$ 2 processes, including $b\bar{b}$).

The following kinematical variables are the most useful ones
for the {\small SM} backgrounds suppression:
lepton $p_T^l$, jet $E_T^j$, $E_T^{miss}$,
scalar transverse energy sum
$E_T^{sum}= \sum p_T^l + \sum E_T^j + E_T^{miss}$ and $Circularity$.
Depending on the {\small mSUGRA} domain under study and
the final state topology, the cut values
are optimized and are typically: $p_T^l > 10 \div 50$ GeV,
$E_T^j > 50 \div 250$ GeV, $E_T^{miss} > 100 \div 500$ GeV,
$E_T^{sum} > 500 \div 1200$ GeV and $C > 0.1$.
Fig.2 shows the 5$\sigma$ $\tilde{g}$/$\tilde{q}$ discovery contours
in various final states with at least two jets, $E_T^{miss} > 100$ GeV
and with one lepton ($1l$),
two leptons of opposite sign ($2l$ OS), two leptons of same sign
($2l$ SS), etc. \cite{sal2}. Clearly, at an integrated luminosity of
$L_{int} = 10^5$ pb$^{-1}$
the gluino/squark masses up to $2 \div 2.5$ TeV
can be probed. The corresponding reach in the $\tilde{\chi}_1^{0}$
mass is $\sim$350 GeV. This result is of particular importance
as the $\tilde{\chi}_1^{0}$ could be a good candidate for
cold dark matter of the Universe. The upper
limit on neutralino relic density corresponds to $\Omega h^2=1$
contour (see Fig.2) \cite{baer2}, which is fully contained
in the explorable domain.

\vspace*{0.7cm}

\begin{figure}[th]
\noindent
\begin{minipage}[b]{.48\linewidth}
\hspace*{-2.cm}
\centering\epsfig{figure=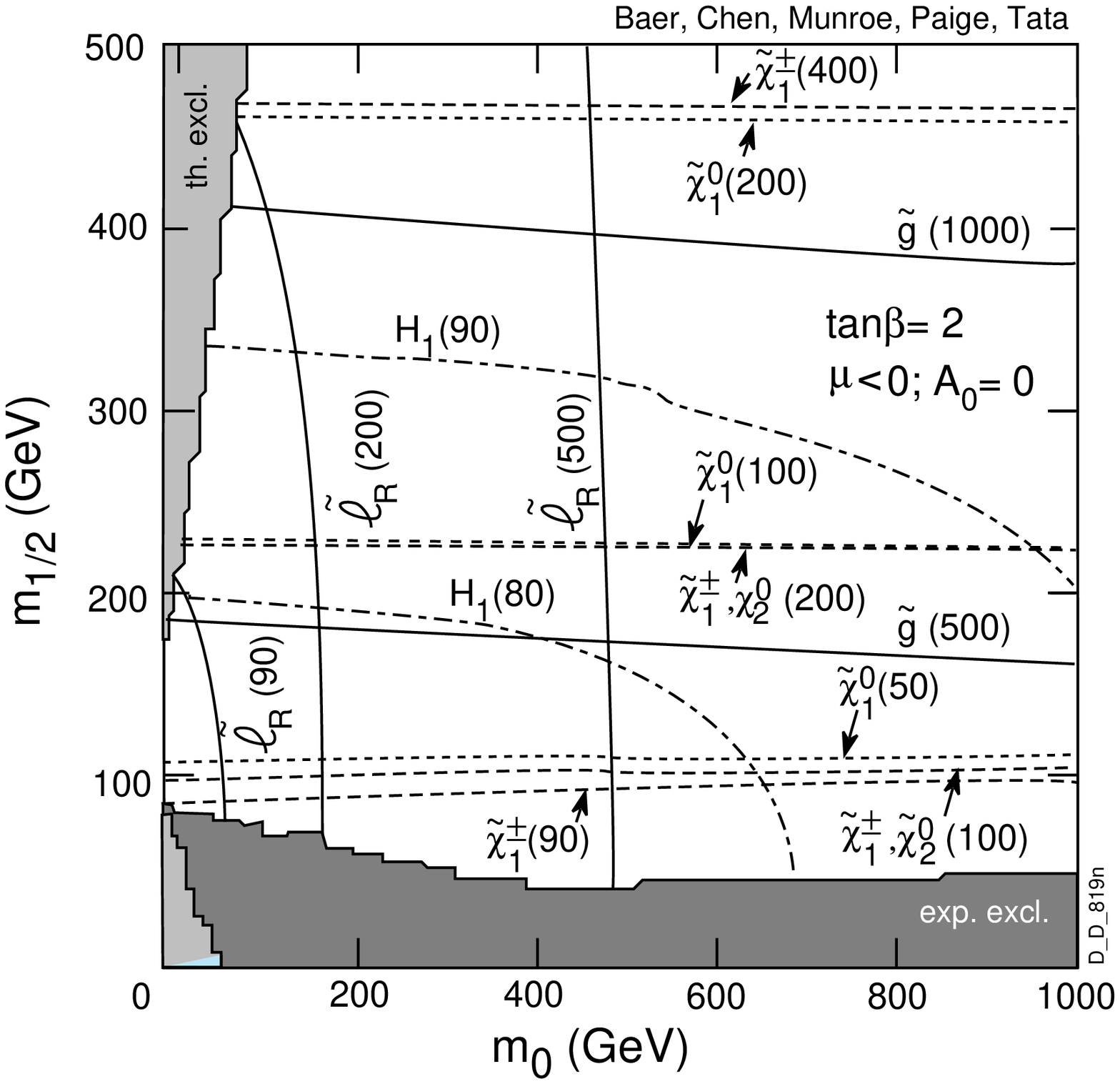,width=8.5cm,height=9.cm}
\vspace{-2.8cm}
\caption{ {\small Sparticles isomass contours in ($m_{0}$, $m_{1/2}$)
parameter space of mSUGRA.} }
\end{minipage}\hfill
\begin{minipage}[b]{.48\linewidth}
\hspace*{-2.7cm}
\centering\epsfig{figure=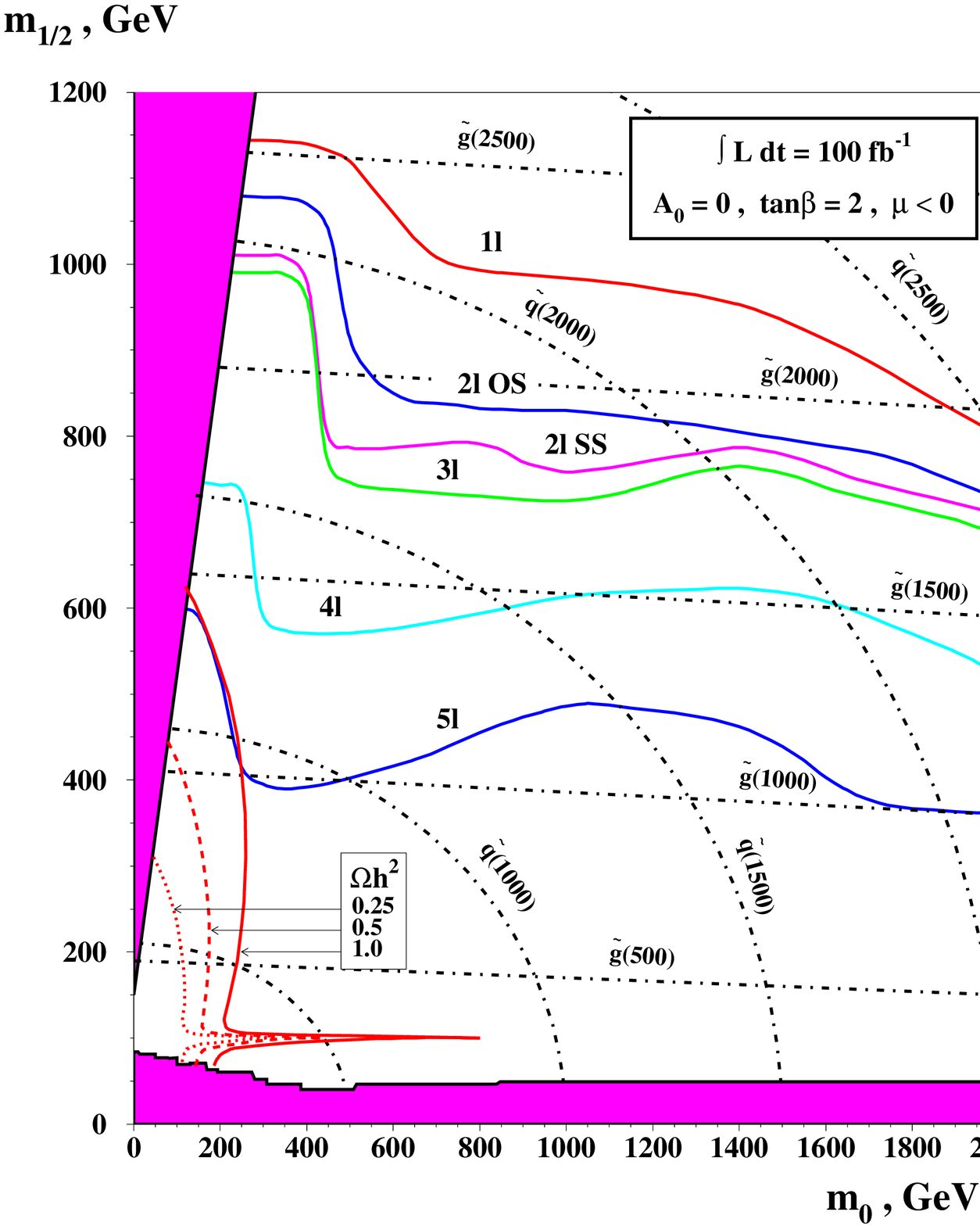,width=7.7cm,height=6.7cm}
\vspace{-0.4cm}
\caption{ {\small Explorable domain of ($m_{0}$, $m_{1/2}$)
parameter space in $n \cdot l + m \cdot jets + E_T^{miss}$ final states
at $L_{int}=10^5$ pb$^{-1}$.} }
\end{minipage}
\vspace*{-0.3cm}
\end{figure}

\section{Slepton Production}

Slepton pairs produced in a Drell-Yan process and decaying
leptonically lead to the final states characterized by
two hard, same-flavor, opposite-sign isolated leptons,
$E_T^{miss}$ and no jet activity, except from the initial
state radiation.
The issue here is to understand and
keep under control the {\small SM} and internal {\small SUSY} backgrounds
in the many processes involved.
The following {\small SM} processes may contribute to 2 leptons final
state:
WW, WZ, $t \bar{t}$, W$tb$, $\tau\tau$, $b\bar{b}$.
A typical set of cuts allowing to extract the signal is:
2 isolated leptons of $p_T^l > 30$ GeV, $E_T^{miss} > 80$ GeV,
veto on jets with $E_T^j > 30$ GeV in $|\eta| < 4$, relative
azimuthal angle between the leptons and $E_T^{miss} > 160^{\circ}$
\cite{vigi}.
Depending on {\small mSUGRA} domain under study the cuts are optimized.
The dominant backgrounds are: reducible $t \bar{t}$
and irreducible WW, $\tilde{\chi}_1^{\pm}\tilde{\chi}_1^{\mp}$.
Other {\small SUSY} backgrounds are largely reduced by
vetoing the central jets. Fig.3 shows {\small mSUGRA} points for
which the slepton signal visibility has been investigated
in the $2l + no \ jets + E_T^{miss}$ final states along with
a 5$\sigma$ significance contour for an integrated luminosity
of $L_{int} = 10^5$ pb$^{-1}$. The slepton mass domain that
can be explored extends up to $\sim$400 GeV.

\section{Chargino-Neutralino Pair Production}

There are 21 different reactions
(8$\tilde{\chi}_{\imath}^{\pm} \tilde{\chi}_{\jmath}^{0}$,
3$\tilde{\chi}_{\imath}^{\pm} \tilde{\chi}_{\jmath}^{\mp}$ and
10$\tilde{\chi}_{\imath}^{0} \tilde{\chi}_{\jmath}^{0};$
$\imath= 1, \ 2;$ $\jmath = 1 \div 4$) for
chargino-neutralino pair production via Drell-Yan
processes, among which \chha \chnb \ production has the largest
cross-section. The easiest way to extract the signal is
to exploit \chha \chnb \ leptonic decays
resulting to the $3l + no \ jets + E_T^{miss}$ final states.
The {\small SM} backgrounds considered are:
WZ, ZZ, $t\bar{t}$, W$tb$, Z$b\bar{b}$, $b\bar{b}$.
The {\small SUSY} processes, that may lead to 3 leptons in final states
are also taken into account:
strong production ($\tilde{g} \tilde{g}, \tilde{g} \tilde{q},
\tilde{q} \tilde{q}$), associated production ($\tilde{g} \tilde{\chi},
\tilde{q} \tilde{\chi}$), chargino-neutralino pair production
($\tilde{\chi} \tilde{\chi}$, other than \chha \chnb) and
slepton pair production ($\tilde{l} \tilde{l}, \tilde{l} \tilde{\nu},
\tilde{\nu} \tilde{\nu}$). The {\small SUSY} background
is dominated by strong production, but the jet veto requirement is very
efficient in reducing \g/\q \ events which in their cascade decays
produce many jets. The typical set of cuts to extract the signal is:
3 isolated leptons with $p_T^l > 15$ GeV; veto on jets with
$E_T^j > 25$ GeV in $|\eta| < 3.5$; a Z mass window cut
$M_Z - 10$ GeV $ < M_{l^{+}l^{-}} < M_Z + 10 $ GeV,
which significantly reduces the dominant {\small SM} WZ contribution.

The signal observability contours at various integrated
luminosities are shown in Fig.4 \cite{charg_neut}.
At low integrated luminosity,
$L_{int} = 10^4$ pb$^{-1}$, the direct production of \chha\chnb \
can be extracted up to $m_{1/2} \lsim 150$ GeV for all $m_0$.
A further increase of
luminosity up to $L_{int} = 10^5$ pb$^{-1}$ extends the explorable
region only by about $10 \div 20$ GeV for $m_0 \gsim 120$ GeV
because of $\tilde{\chi}_2^{0} \rightarrow
l^+l^-\tilde{\chi}_1^{0}$ 3-body decays
are overtaken by "spoiler" modes
$\tilde{\chi}_2^{0} \rightarrow$ h$\tilde{\chi}_1^{0}$,
Z$\tilde{\chi}_1^{0}$, which are now kinematically allowed.
However, for $m_0 \lsim 120$ GeV the
gain in parameter space is much larger, up to $m_{1/2} \lsim$ 420 GeV,
due to
$\tilde{\chi}_2^{0} \rightarrow l^+ \tilde{l}_{R,L}^-
\rightarrow l^+l^-\tilde{\chi}_1^{0}$ 2-body decays.

\vspace*{1.cm} 

\begin{figure}[th]
\noindent
\begin{minipage}[b]{.48\linewidth}
\hspace*{-2.cm}
\centering\epsfig{figure=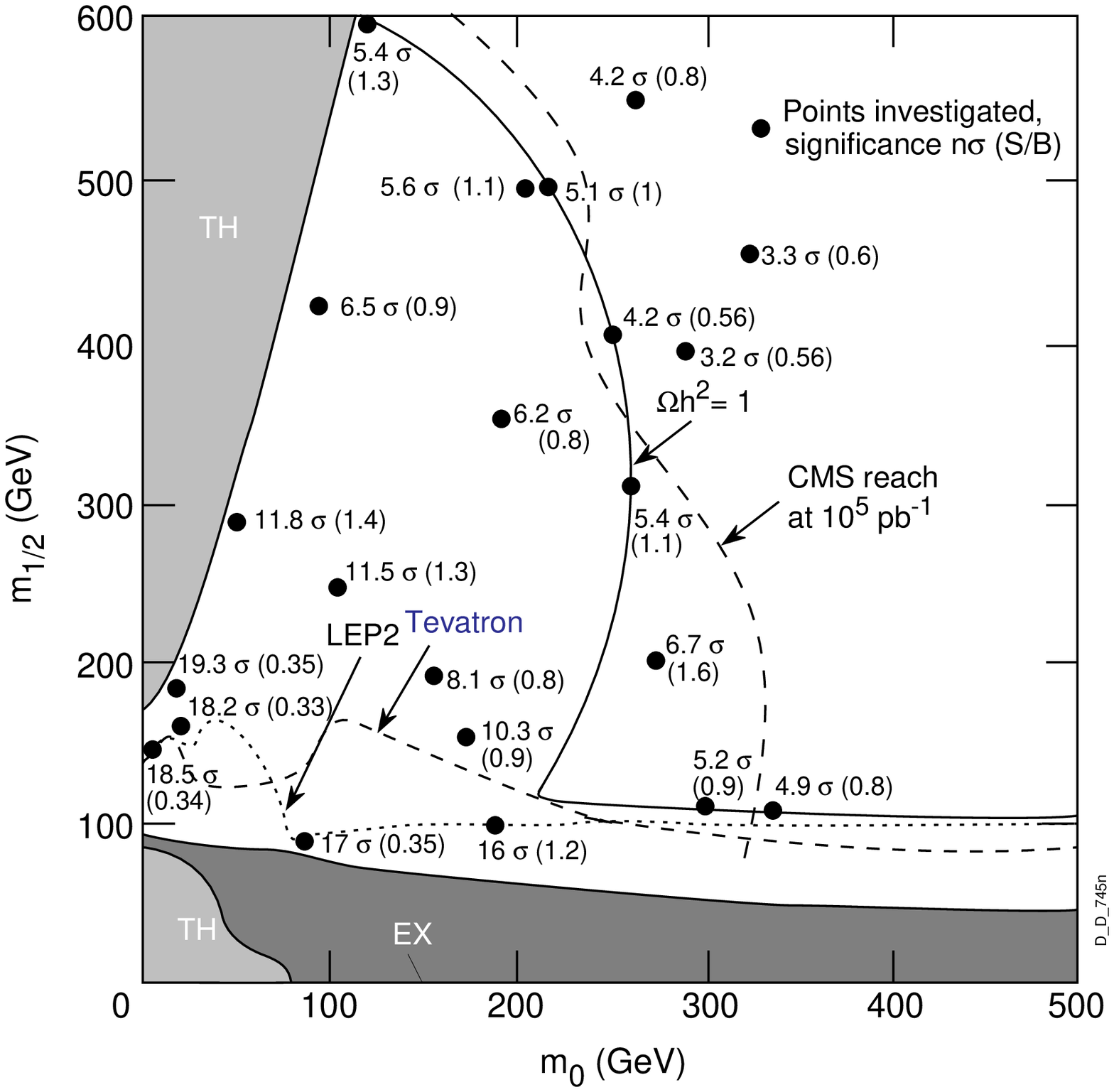,width=8.5cm,height=8.5cm}
\vspace{-2.9cm}
\caption{ {\small Explorable domain of ($m_{0}$, $m_{1/2}$)
parameter space in slepton searches.} }
\end{minipage}\hfill
\begin{minipage}[b]{.48\linewidth}
\hspace*{-2.cm}
\centering\epsfig{figure=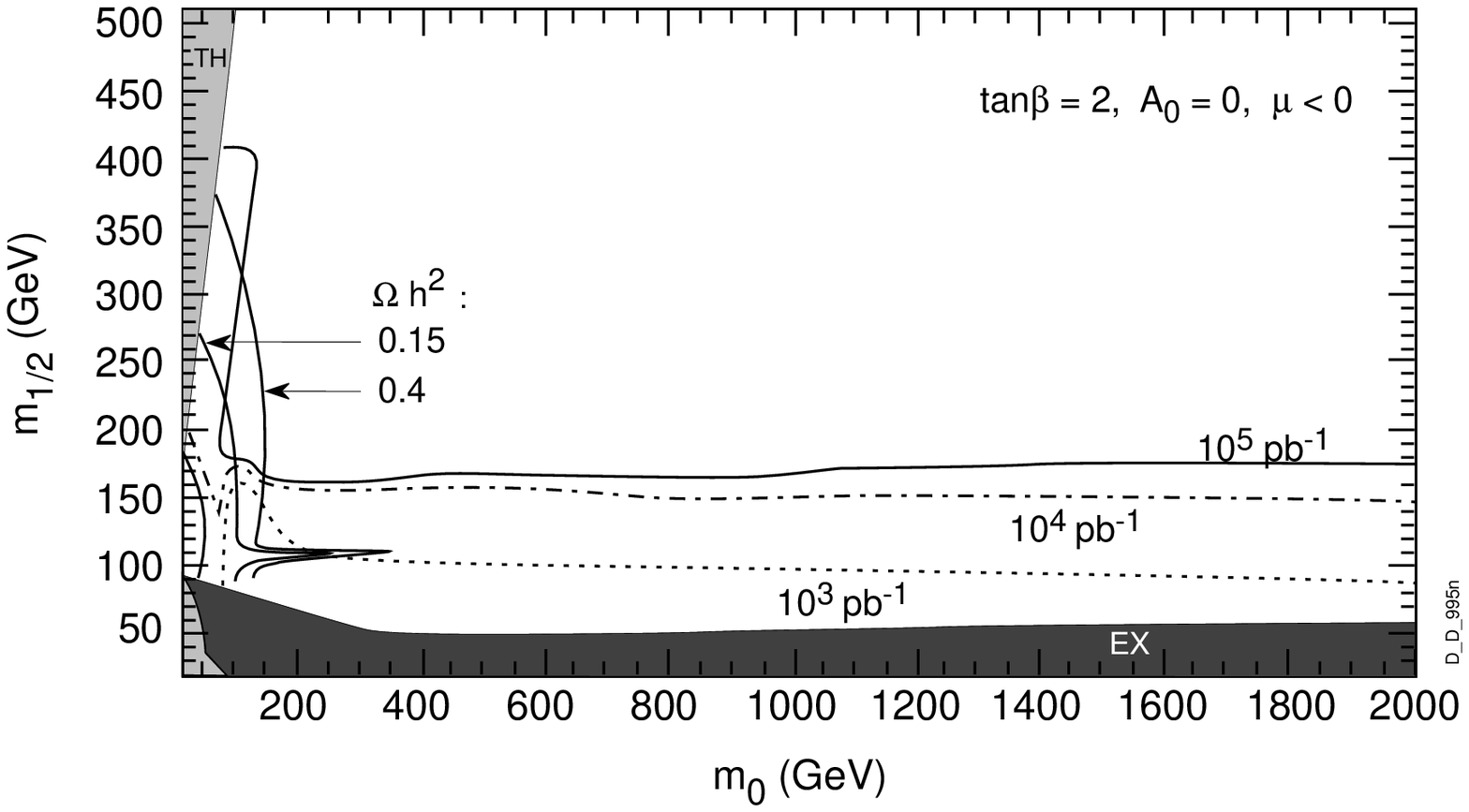,width=9.cm,height=12.cm}
\vspace{-8.cm}
\caption{ {\small 5$\sigma$ significance contours
at various luminosities for Chargino-Neutralino direct production.} }
\end{minipage}
\vspace*{-0.3cm}
\end{figure}

\section{Search for the Next-to-Lightest Neutralino}

The kinematical feature of $\tilde{\chi}^{0}_{2}$
leptonic decays which can be exploited
is the $l^{+}l^{-}$ invariant mass spectrum
with its characteristic edge, as shown in Fig.5 for a
representative {\small mSUGRA} point. In cascade decays of
gluinos and squarks the production of
$\tilde{\chi}^{0}_{2}$ is so abundant, that an "edge" in the dilepton mass
spectrum can be observed in a significant part of ($m_0$, $m_{1/2}$)
plane. Fig.6 shows the cross-section times branching ratio into leptons
for $\tilde{\chi}^{0}_{2}$ inclusive production.
The SM background can be easily suppressed, e.g. requiring a third
lepton or/and $E_T^{miss}$, and with $L_{int} = 10^5$ pb$^{-1}$
a detectable edge is seen as long as $\sigma \cdot$B $\gsim 10^{-2}$ pb
\cite{chi2}.

There are domains of parameter space when both 2- and 3-body decays
co-exist (or where both modes,
$\tilde{\chi}_2^{0} \rightarrow l^+ \tilde{l}_{R}^-$ and
$l^+ \tilde{l}_{L}^-$ are present). For example,
in Fig.7 the first edge
at $M^{max}_{l^{+}l^{-}} = 52$ GeV
corresponds to $\tilde{\chi}_2^{0} \rightarrow l^+ \tilde{l}_{R}^-$
decays, while the second edge at $M^{max}_{l^{+}l^{-}} = 69$ GeV to
$\tilde{\chi}_2^{0} \rightarrow l^+l^-\tilde{\chi}_1^{0}$ ones.
In 3-body decays,
the kinematical upper limit of the dilepton mass spectrum is
$M^{max}_{l^{+}l^{-}} = M_{\tilde{\chi}^{0}_{2}} -
M_{\tilde{\chi}^{0}_{1}}
\approx M_{\tilde{\chi}^{0}_{1}}$,
as in {\small mSUGRA}
$M_{\tilde{\chi}^{0}_{2}} \approx 2 M_{\tilde{\chi}^{0}_{1}}$,
thus providing the determination of the
$\tilde{\chi}^{0}_{1}$ mass.
Whether an edge corresponds to a 2- or 3-body decays can be
deduced from the decay kinematics, e.g. lepton $p_T$-asymmetry
$A = (p_T^{max} - p_T^{min})/(p_T^{max} + p_T^{min})$.
This is illustrated in Fig.8, where several dotted histograms
are obtained assuming, that the first edge in Fig.7 is due to the 3-body
decays with $M_{\tilde{\chi}^{0}_{2}} - M_{\tilde{\chi}^{0}_{1}} = 52$
GeV.

\vspace{0.6cm}

\begin{figure}[th]
\noindent
\begin{minipage}[b]{.48\linewidth}
\hspace*{-1.8cm}
\centering\epsfig{figure=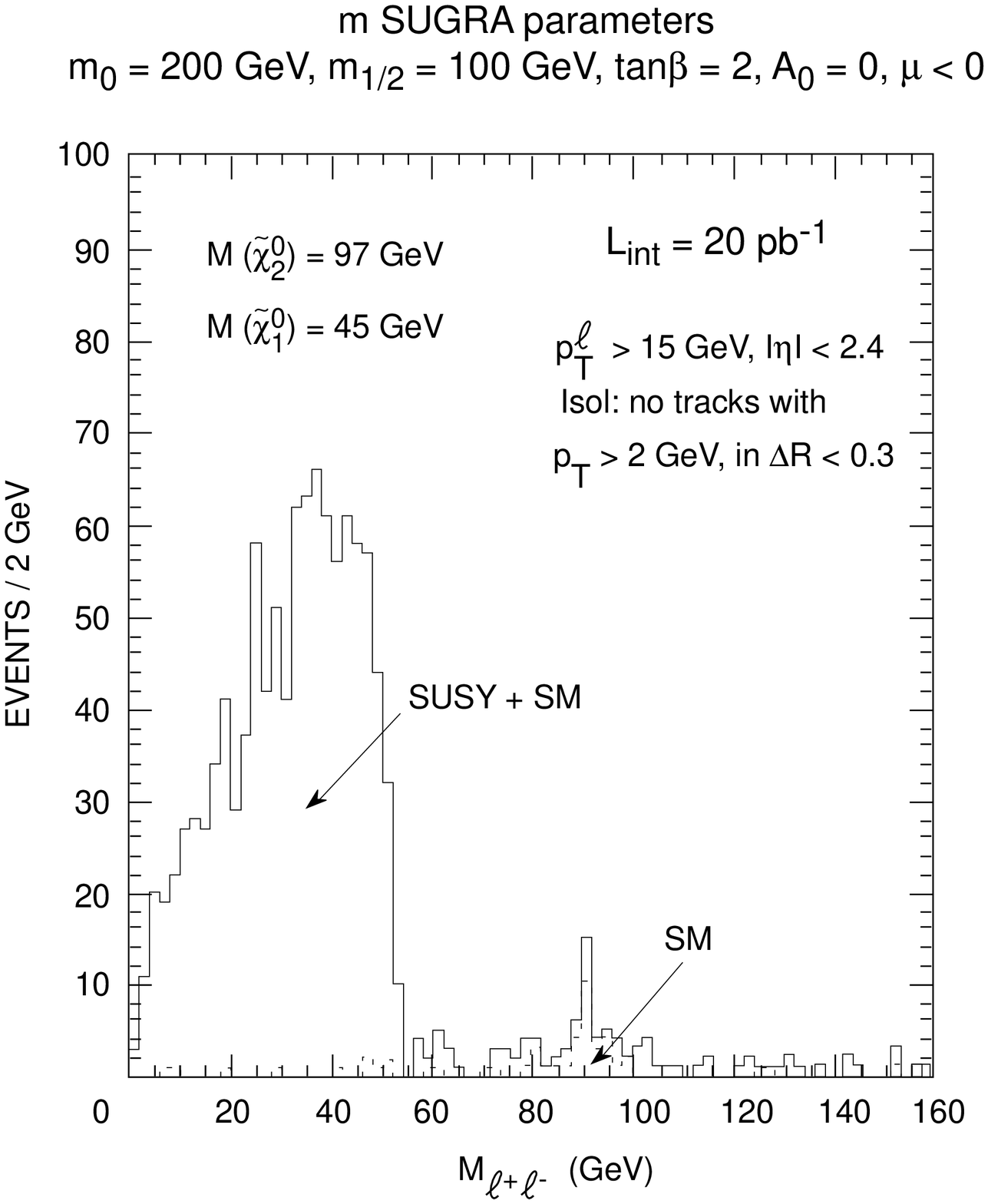,width=8.3cm,height=8.cm}
\vspace{-0.8cm}
\caption{ {\small $l^+l^-$ mass spectrum in 3-lepton final states.} }
\end{minipage}\hfill
\begin{minipage}[b]{.48\linewidth}
\hspace*{-1.8cm}
\centering\epsfig{figure=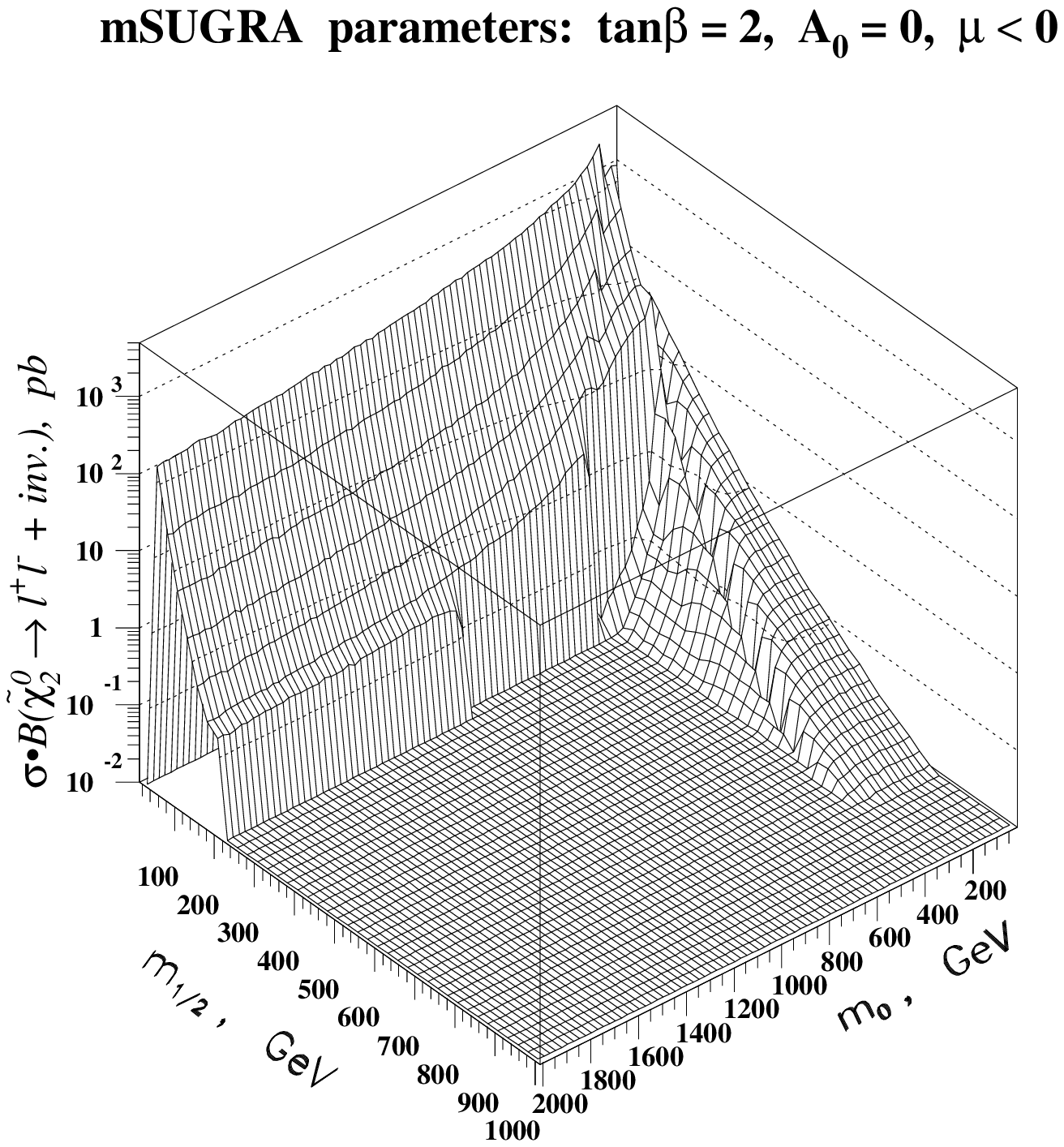,width=11.5cm,height=12.cm}
\vspace{-5.6cm}
\caption{ {\small $\tilde{\chi}_{2}^{0}$ inclusive production
cross-section times branching ratio into leptons.} }
\end{minipage}
\vspace*{-0.1cm}
\end{figure}

\vspace{0.5cm}

\begin{figure}[th]
\noindent
\begin{minipage}[b]{.48\linewidth}
\hspace*{-1.6cm}
\centering\epsfig{figure=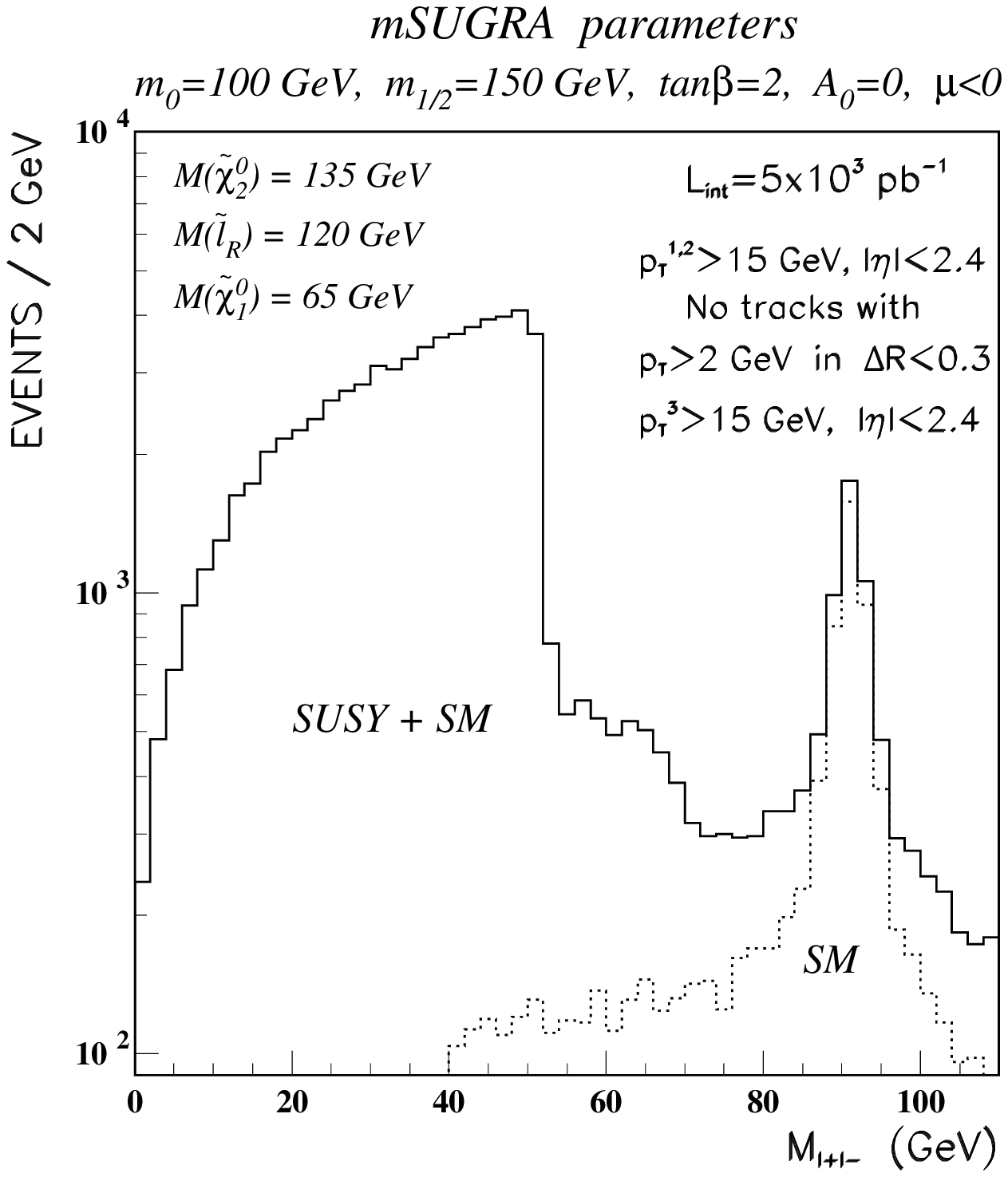,width=11.cm,height=11.5cm}
\vspace{-5.2cm}
\caption{ {\small $l^+l^-$ mass spectrum in 3-lepton final states.} }
\end{minipage}\hfill
\begin{minipage}[b]{.48\linewidth}
\hspace*{-1.6cm}
\centering\epsfig{figure=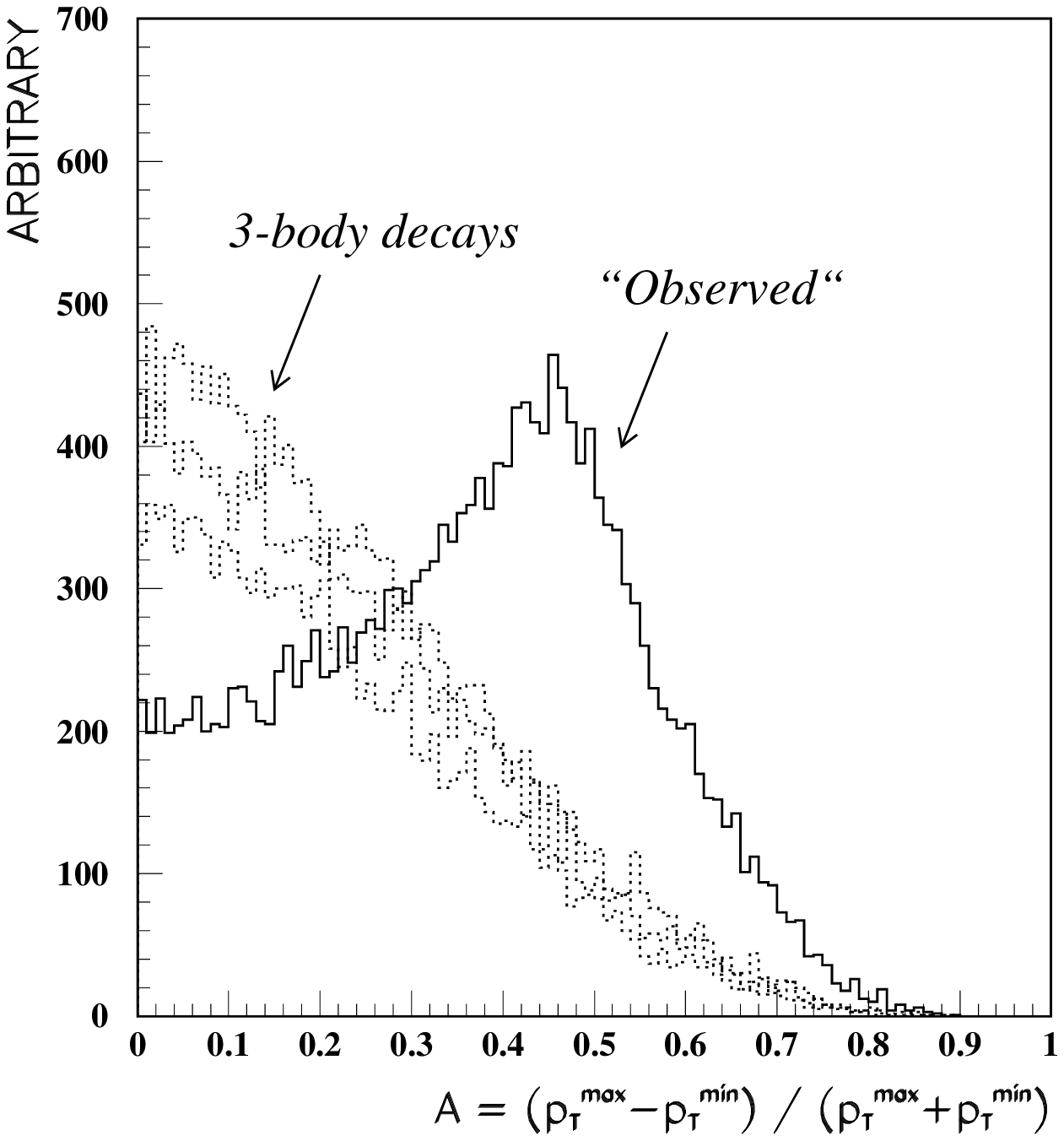,width=11.cm,height=11.5cm}
\vspace{-5.1cm}
\caption{ {\small Leptons $p_T$ asymmetry distributions in
3- and 2-body decays of $\tilde{\chi}_2^{0}$.} }
\end{minipage}
\vspace*{-0.3cm}
\end{figure}

In 2-body decays the kinematical upper limit of the dilepton
mass spectrum is
$M^{max}_{l^{+}l^{-}} = \sqrt{
(M^{2}_{\tilde{\chi}^{0}_{2}} - M^{2}_{\tilde{l}})
(M^{2}_{\tilde{l}} - M^{2}_{\tilde{\chi}^{0}_{1}}) } /
M_{\tilde{l}}$,
i.e. it is sensitive also to the slepton mass,
which can be determined in the following way:
i) assume $M_{\tilde{\chi}^{0}_{2}} = 2 M_{\tilde{\chi}^{0}_{1}}$,
ii) generate samples of $\tilde{\chi}^{0}_{2}$ 2-body sequential
decays for various $M_{\tilde{\chi}^{0}_{1}}$
($M_{\tilde{l}}$); note, that the slepton mass is constrained
to provide the "observed" position of an edge, iii) ambiguity
is then resolved statistically, by means of, e.g.
a $\chi^2$-test of the shape of the lepton $p_T$-asymmetry distributions.
For the {\small mSUGRA} point ($m_0=100$ GeV, $m_{1/2}=150$ GeV)
this procedure provides the following precisions on masses:
$\delta M_{\tilde{\chi}^{0}_{1}} \lsim$ 5 GeV,
$\delta M_{\tilde{l}} \lsim$ 10 GeV.
The use of both edges yields
$\delta M_{\tilde{\chi}^{0}_{1},\tilde{l}} \lsim$ 1 GeV.

Note, that the Z peak seen in Fig.5,~7 serves
as an overall calibration signal; it allows to control
the mass scale as well as the production cross-section. 

\section{Conclusions}

At {\small LHC/CMS} {\small SUSY} will reveal itself easily by an
excess of events
over {\small SM} expectations in a number of characteristic signatures:

\begin{itemize}

\vspace{-0.3cm}

\item
The $n \cdot l + m \cdot jets + E_T^{miss}$ \
final states
provide the maximal reach for {\small SUSY}
via production of the strongly interacting sparticles.
$\tilde{g}$ and $\tilde{q}$ masses can be probed up to
$\sim$2.5 TeV.
The explorable $\tilde{\chi}_1^{0}$
mass extends up to $\sim$350 GeV.

\vspace{-0.3cm}

\item
The $2l + no \ jets + E_T^{miss}$ \
final states give access to slepton pair production.
The explorable slepton
mass range would extend up to $\sim$400 GeV.

\vspace{-0.3cm}

\item
The $3l + no \ jets + E_T^{miss}$ \
is the final state in which
direct $\tilde{\chi}^{\pm}_{1}\tilde{\chi}_2^{0}$ production
in a Drell-Yan process should be looked for.

\vspace{-0.3cm}

\item
The $l^+l^- + \ l^\pm/E_T^{miss}$ \
final states, with relatively
modest demands on detector performance,
may well be the first channel in which {\small SUSY} would reveal
itself through $\tilde{\chi}_2^{0}$ inclusive production with subsequent
decays -- directly or via sleptons -- into
$l^+l^-\tilde{\chi}_1^{0}$. Characteristic dilepton
invariant mass spectrum
with a spectacular edge could provide a good handle
to determine the sparticle masses.

\end{itemize}

\vspace{-0.3cm}

What can be measured depends on a scenario
Nature has chosen, but if {\small SUSY} is of relevance at {\small EW}
scale it could hardly escape detection at {\small LHC/CMS}.

\vspace{5mm}

{\bf {Acknowledgments}}

\vspace{2mm}

I am thankful to my colleagues from the {\small CMS/SUSY}
working group led by Dr. Daniel Denegri. Special thanks
to the organizers of the Moriond'97 for the support.

\end{document}